\documentclass[fleqn,usenatbib]{mnras}

\usepackage{newtxtext,newtxmath}

\usepackage[T1]{fontenc}
\usepackage{ae,aecompl}

\usepackage{siunitx}
\usepackage{placeins}

\usepackage{graphicx}	
\usepackage{amsmath}	
\usepackage[dvipsnames]{xcolor}

\newcommand{\tripleseven}{O I $7774\,\si{\angstrom}$ }
\newcommand{\halpha}{H$\alpha$}
\newcommand{\hbeta}{H$\beta$}
\newcommand{\novacar}{V906 Car}
\newcommand{\mgab}{YZ Reticuli}
\newcommand{\novasgr}{V5668 Sgr}
\newcommand{\pnvj}{V6568 Sgr}

\newcommand{\alma}{ALMA}
\newcommand{\swift}{\textit{Swift}}

\newcommand{\asassn}{ASAS-SN}

\newcommand{\infrared}{infrared}
\newcommand{\xray}{X-ray}
\newcommand{\xrays}{X-rays}
\newcommand{\pcygni}{P-Cygni}
\newcommand{\Aquila}{Aquila}
\newcommand{\exposure}[1]{$#1\,$s}
\newcommand{\codeendeavour}{\textsc{endeavour}}
\newcommand{\poirot}{\textsc{poirot}}

\definecolor{orange2}{RGB}{226, 117, 32}

\newcommand{\about}{$\sim$}
\newcommand{\GJW}{Global Jet Watch}

\newcommand{\kms}[1]{$#1\,\si{\km\per\second}$}

\newcommand{\rne}{recurrent novae}
\newcommand{\cn}{classical nova}
\newcommand{\cne}{classical novae}

\newcommand{\Cn}{Classical nova}

\newcommand{\Cne}{Classical novae}
\newcommand{\HeN}{\ion{He}{}/\ion{N}{}}
\newcommand{\fesixtyeightyfive}{[\ion{Fe}{vii}] 6085\,Å}
\newcommand{\oxygensixtythreehundred}{[\ion{O}{i}] 6300\,Å}
\newcommand{\nitrogensixfiveeightfour}{[\ion{N}{ii}] 6584\,Å}
\newcommand{\feii}{\ion{Fe}{ii}}
\newcommand{\angstrom}[1]{$#1\,\si{\angstrom}$}
\newcommand{\days}[1]{$+#1\,$d}
\newcommand{\squarebrackets}[1]{\phantom{\null}$^{\textrm{[#1]}}$}
\newcommand{\jd}[1]{$#1\,$JD}
\DeclareSIUnit \parsec {pc}
\newcommand{\gaussian}{Gaussian}
\newcommand{\gaussians}{Gaussians}
\newcommand{\solarmasses}[1]{$#1\,M_{\odot}$}

\title[The onset of jets in classical novae]{The onset of jets in classical novae}

\author[D. McLoughlin et al.]{
Dominic McLoughlin,$^{1}$\thanks{E-mail: dominic.mcloughlin@physics.ox.ac.uk (DM)}
Katherine M. Blundell,$^{1}$
Steven Lee, $^{2, 3}$
Chris McCowage $^{2, 3}$
\\
$^{1}$Department of Physics, University of Oxford, Keble Rd, Oxford OX1 3RH, United Kingdom\\
$^{2}$Research School of Astronomy and Astrophysics, Australian National University, Canberra, ACT 2611\\
$^{3}$Anglo-Australian Telescope, Coonabarabran NSW 2357, Australia\\
}

\date{Accepted 2021 May 05. Received 2021 April 29; in original form 2021 March 08}

\pubyear{2021}

\begin{document}
\label{firstpage}
\pagerange{\pageref{firstpage}--\pageref{lastpage}}
\maketitle

\begin{abstract}
We present two further classical novae, V906 Car and V5668 Sgr, that show jets and accretion disc spectral signatures in their H-alpha complexes throughout the first 1000 days following their eruptions. From extensive densely time-sampled spectroscopy, we measure the appearance of the first high-velocity absorption component in V906 Car, and the duration of the commencement of the main H-alpha emission. We constrain the time taken for V5668 Sgr to transition to the nebular phase using [N II] 6584Å.  We find these timings to be consistent with the jet and accretion disc model for explaining optical spectral line profile changes in classical novae, and discuss the implications of this model for enrichment of the interstellar medium.
\end{abstract}




\section{Introduction} \label{sec:introduction}
\Cne\ are explosive events which occur on the surface of white dwarfs found in close binaries. Hydrogen accretes onto the degenerate surface, and heats up until it reaches temperatures of around $10^8\,$K, when it ignites and thermonuclear runaway occurs \citep{GallagherStarrfield1978}. This ejects a shell of the accreted hydrogen, whose mass is dependent on that of the underlying white dwarf. The exact nature of the outflows from such an eruption are not well known, and there are competing theories about interactions between the potentially complex ejecta and the immediate surroundings \citep{Arai2016,Williams2010}. 

One particular topic about which not much is known is the accretion disc and its survival in the immediate aftermath of a \cn. While simulations suggest that such a disc might be destroyed by the eruption \citep{Figueira2018}, observational evidence indicates that discs may survive in some form \citep{Skillman1997,Hounsell2011,Hachisu2012}. This paper follows \citet{McLoughlin2021} which presented, from an analysis of time-resolved spectroscopy, evidence for the persistence of both jets and an accretion disc in the weeks and months that followed the eruption of \cn\ \mgab. A key significance of the surviving accretion disc is that a disc is considered a necessary (but not sufficient) condition for the launching of jets \citep{Gouveia2005,Coppejans2020}.  In the present paper, we add two new examples, \novasgr\ and \novacar, which also seem to display the spectral signature of accretion discs from early times after the eruption. We also detect jets in both of these novae, and are able to time the onset of their becoming spectrally significant in \halpha\ observations.

Astrophysical jets are a ubiquitous phenomenon arising from central masses having a huge dynamic range of mass, from supermassive black holes giving rise to synchrotron-emitting jets in quasars, to stellar-mass black holes in microquasars giving jets that are both line-emitting and synchrotron, to young stellar objects having line-emitting jets. Recently-launched line-emitting jets from \cne\ therefore afford a new opportunity to study the commencement of jet launch in lower-mass systems.

Objects classified as \cne\ display a rich variety of behaviours, yet there are many phases which most novae appear to exhibit.  We use densely time-sampled spectroscopic observations of three \cne\ to tightly constrain the durations of certain transitions between phases for these particular objects. 

In Section \ref{sec:observations_novae}, we explain the data collected for each target. In Section \ref{sec:novacar_timing_phases}, we quantify the duration of certain phase changes for \novacar. In Section \ref{sec:novasgr_timing_phases}, we perform a similar set of analyses for \novasgr, which also reveal the spectral signature of jets and an accretion disc. In Section \ref{sec:novae_overview}, we discuss the jets and disc of \novacar\ and the overall picture of \cn\ evolution in light of this model, and consider the impact nova jets may have on enrichment of the interstellar medium. In Section \ref{sec:conclusions}, we conclude by recommending new observations to target direct observations of such jets.

\subsection{Classification schemes for \cne}
Over the years, there have been several efforts to categorise different aspects of the hugely varied range of behaviours exhibited by \cne\ in spectroscopic, photometric and polarimetric data. A thorough account of our modern understanding of the phases typically exhibited by \cne\ is given in section 4 of \citet{DellaValle2020}, and so here we give just a brief summary of a relevant subset of this material.

In terms of spectral characteristics, \citet{Williams1991,Williams1992,Williams1994} performed a deep analysis of optical spectra for a set of 13 novae, and found that they separate into two spectral classes --- \feii\ and \HeN, a designation known as the Tololo scheme. The \feii\ spectral type is typified by having many low-excitation \feii\ emission lines, along with \pcygni\ profiles shortly after eruption in low-ionisation heavy elements, and these spectra are thought to be formed in a circumbinary reservoir, fed by the wind off the secondary. \HeN\ novae on the other hand, show strong flat-topped emission lines of helium and neon, hinting at their formation in the white dwarf ejecta. They also found that some \cne\ exhibit a ``hybrid'' evolution, changing between the two types.

\begin{figure*}
	\includegraphics[width=\textwidth]{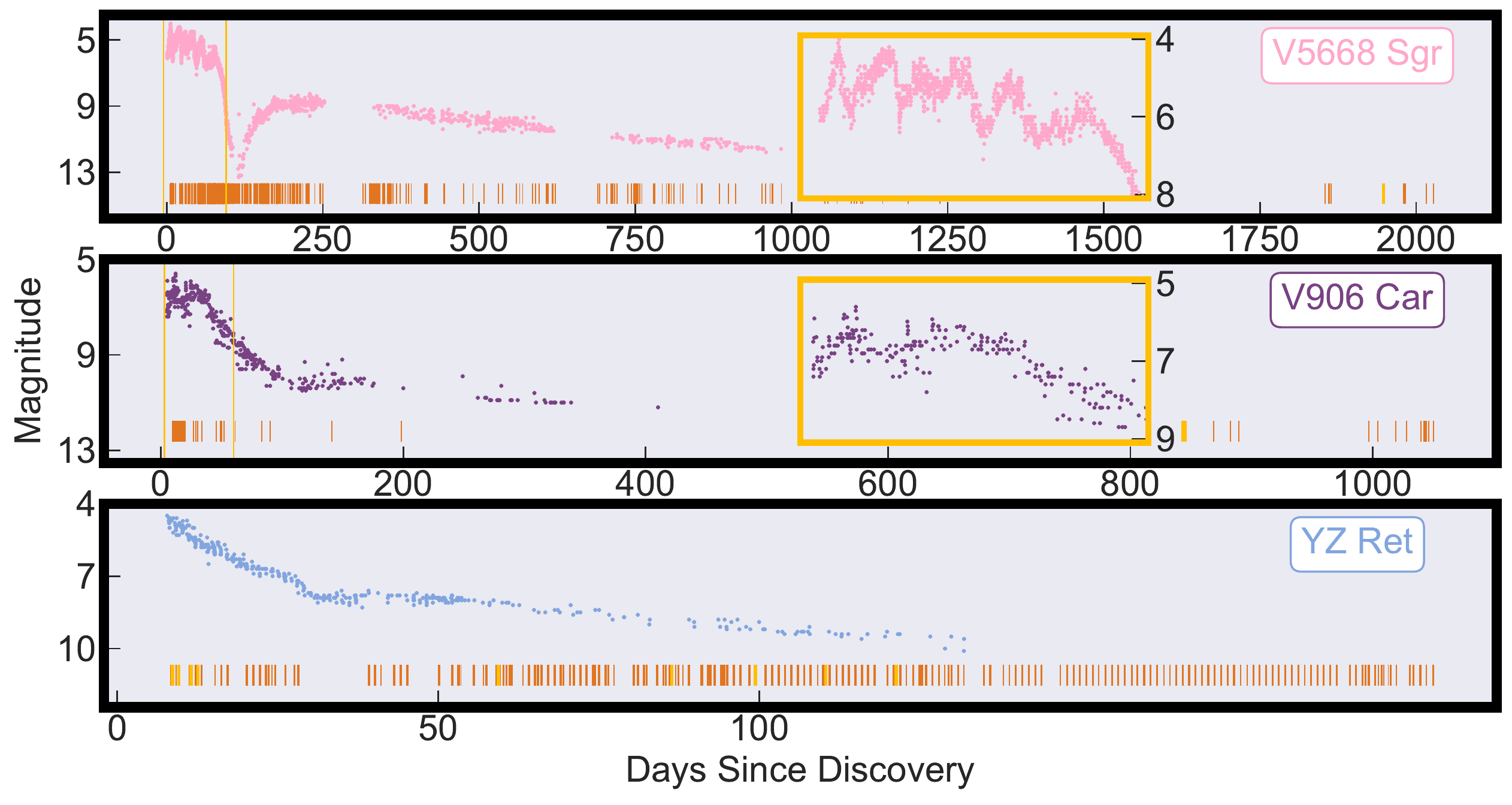}
    \caption{AAVSO visual light curves for the three \cne\ studied in this paper in chronological order of eruption. \novasgr\ shows a deep dust-dip phase between \days{80} and \days{200}, with some oscillations immediately following the maximum lasting until at least the start of the dust-dip. \novacar\ has initial flares, shown to be correlated with \xray\ emission \citep{Aydi2020}. \mgab\ exhibits a plateau between \days{31} and \days{62}. The gaps in the \novasgr\ light curve (commencing at roughly \days{250} and \days{620}) correspond to when it was a daytime object. The epochs at which we have Aquila spectra are marked with orange ticks in the main panels, with LHires observations marked in gold. The gold boxes in the upper two panels show close-ups of flares present in the early stages of those eruptions.}
    \label{fig:novae_light_curves}
\end{figure*}

Light curves of \cne\ can be organised into several broad types based on features they exhibit \citep{Strope2010}. For instance, a ``dust dip'' light curve (like that of \novasgr, shown in the upper panel of Figure \ref{fig:novae_light_curves}) indicates condensation of dust in the nova ejecta, leading to a months-long dip in optical brightness and a corresponding increase in the infrared, while a ``smooth'' light curve follows a simple broken power law of decline, due to radiative transfer in the shell ejecta.

\Cn\ light curves also have speed classes, with typical designations based on the time to decline by two optical magnitudes from maximum (known as $t_2$), although this can be difficult to assign in cases when the maximum is not well-sampled or the decline is not smooth. It is thought that the faster a \cn\ fades, the more massive the white dwarf is, and the more rapidly it will recur \citep{Kato2001,Shara2018}.

Studies of \cne\ in the \xray\ have revealed a clear distinction between hard and soft \xray\ radiation, with the former being associated with shocks in outflows and the latter with the surface burning. \citet{Schwarz2011} performed a population study of the \xray\ behaviour of \cne, and determined that systems with a faster speed class show early hard \xrays, which are not present for slow novae. 

\subsection{Jets and shell ejecta in novae} \label{sec:jets_and_shell_ejecta}
The previous paper in this series \citep{McLoughlin2021} studied the \halpha\ complex of \cn\ \mgab, and found it to be well-fit by the same five \gaussian\ emission components throughout the many months following its July 2020 eruption. Four of the five components naturally divide into two pairs, with each pair consisting of a red-shifted and a blue-shifted component; in addition there was a fifth component which did not appear to be dynamically significant.

\citet{McLoughlin2021} made the case for oppositely-directed outflows in \mgab\ i.e.\ jets.  We also presented evidence that the slower-moving pair might arise from an accretion disc, which would provide a means for launching the jets. For each of the two pairs, some intriguing patterns appeared as can be seen in figures 6 and 7 of \citet{McLoughlin2021}. For example, the radial-velocity separation within each pair increased and decreased repeatedly, because of symmetric, correlated motions of each of its constituent components.

The most natural interpretation of this behaviour, as presented in \citet{McLoughlin2021}, is that the angle with which the components are moving with respect to Earth's line of sight was changing.  This phenomenon, in the case of oppositely-directed jets, is known as precession. Those figures show that the varying separation of the higher radial-velocity pair of lines does not have a simple phase-relationship with that of the lower-speed pair.

Precession in jets is likely to arise because of precession of the accretion disc which launches those jets; this is directly evidenced in the case of the steadily precessing accretion disc and jets of the microquasar SS433 \citep{Blundell2008}. In contrast, a newly-erupted \cn\ will have an unbalanced accretion disc that has not yet equilibrated that is likely to precess as it launches the jets; the phase-relationship between its precession and that of the jets will not be as straightforward to identify.

\citet{McLoughlin2021} further found that a similar model accounted for the emission in \pnvj, a \cn\ which rapidly faded and so did not warrant a major observing campaign. In the present paper we extend that work by examining historical data captured in two major campaigns by the \GJW\ on two previous \cne, \novacar\ and \novasgr, with the jets and disc model found to describe the behaviour of \mgab. 

\citet{McLoughlin2021} proposed that \pnvj\ bridges the gap in white dwarf mass between \mgab\ and the \rne\ (that are already known to produce jets e.g.\ \citep{Sokoloski2008,Rupen2008}), whose rapid recurrence timescales are believed to be due to their masses being higher than non-recurrent \cn\ systems. White dwarf mass is intrinsically linked to their nova speed class, with higher masses leading to faster recurrence rates and shorter decline times \citep{Schwarz2011}. Speed class therefore may relate to the propensity of a nova system to launch jets along the axis of spin of the white dwarf.

\subsection{\Cn\ \novasgr}
\Cn\ \novasgr\ was discovered by John Seach on 2015 March 15.634 UT (= \jd{2457097.134}), which we take as \days{0} for this paper. It is also known as Nova Sgr 2015b, and PNV J18365700$-$2855420. It has (J2000) Right Ascension 18 37 39.9 and Declination $-$29 04 03,  and Galactic coordinates of 005.3211 $-$10.0678. It was classified as being of type \feii\ by \citet{Banerjee2015} using near-\infrared\ observations.

This nova, like the others we study in this paper, benefits from light curves whose data points have been contributed by many amateur observers around the world.   It is remarkable that despite the non-uniformity of technique and instrumentation (visual vs camera) a clear picture of the changing flux density emerges, as exemplified in Figure \ref{fig:novae_light_curves}.  While there is some scatter arising from such heterogeneous collections of data (see for example the analysis in \citet{Baluev2015} and references therein) this does not in any way hinder the scope of this paper. The AAVSO\footnote{\href{https://aavso.org}{aavso.org}} light curve for this \cn\ (shown in the first panel of Figure \ref{fig:novae_light_curves}) showed a prominent dust-dip from approximately \days{80} to \days{200} \citep{Banerjee2015}. It was found to have a white dwarf mass of $0.85\,M_{\odot}$ by \citet{Hachisu2019}, using their scaled light curve method. 

\citet{Gehrz2018} describes in detail the \xray\ story according to \swift\ observations of \novasgr. Their first detection in \xray\ was at \jd{2457191.5}, which corresponds to \days{94}. They derive supersoft source switch-on and switch-off times as $t_{\text{on}} =$\days{168} and $t_{\text{off}} = $\days{240} to \days{340}. They find the mass of the underlying CO white dwarf to be $1.1\,M_{\odot}$.

\citet{Molaro2016} found \novasgr\ to be a significant producer of $^7$\ion{Be}{ii}, and used this to suggest that \cne\ may be the source of all the $^7$\ion{Li}{} in the Milky Way. \citet{Diaz2018} discovered structures in \alma\ imaging of the ejecta of \novasgr\ which were the smallest scale of those found in the remnants of any stellar explosion so far, although the veracity of the details of the extended emission depends on UV-plane coverage of the interferometric \alma\ observations.

\citet{Jack2017} performed a detailed identification of four optical spectra of \novasgr\ and compared their spectroscopy with the light curve to show that certain blue-shifted absorption features moved to slower radial velocities during the multiple rises to maximum (present until \about{\days{80}} in the light curve of \novasgr\ in the upper panel of Figure \ref{fig:novae_light_curves}), and to faster radial velocities during the decline. We briefly expand on this work in Section \ref{sec:novasgr_nitrogen_triplet}.

\subsection{\Cn\ \novacar}
The earliest detection of \cn\ \novacar\ was at \jd{2458193.819444} by the Evryscope-South observatory, according to \cite{Corbett2018}, which we refer to as the discovery date for the system. It was independently discovered after this by \citet{Atel11454} using the All Sky Automated Survey for SuperNovae (hereafter \asassn). In this paper, we reference times as days since commencement of brightening (\jd{2458193.819444}). It has (J2000) Right Ascension 10:36:15.426 and Declination $-$59:35:53.731; its Galactic coordinates are 286.580 $-$01.088.

The mass of the white dwarf in \novacar\ was estimated at $0.71^{+0.23}_{-0.19}\,M_{\odot}$ through an analysis of the progenitor system by \citet{Wee2020} using an accretion disc model to understand the light curve. Their decline times of $t_{2, v} = 26.2\,$d and $t_{3, v} = 33.0\,$d point to \novacar's being a moderately fast \cn, although \citet{Aydi2020a} suggest a value of $t_{2} = 44\,$d. \citet{Sokolovsky2020} determined that this \cn\ occured on a CO white dwarf, and they concluded that the mass must have been low as there was no distinct supersoft phase detected in the \xray. Their thorough analysis of the NuSTAR, Fermi/LAT, XMM–Newton and Swift/XRT \xray\ data revealed that the nova shell was expelled by \days{24}. The white dwarf was found to be of CO-type by \citet{Sokolovsky2020}.

The optical AAVSO light curve shows some initial flares up to around \days{30}, and exhibited weak cusp-like behaviour between \days{100} and \days{150}. \citet{Aydi2020} used correlated flares in the optical and gamma-ray light curves of \novacar\ to show that the emission in both bands is probably powered by the same underlying mechanism, which they took to be fast outflow colliding with slower outflows, and shock-heating it. Observations by \citet{Pavana2020} show \novacar\ to be of the hybrid class of \cne. \citet{Wee2020} associate \novacar\ with the Carina Nebula, and thus assume a distance to the \cn\ of $d = 2.3 \pm 0.5$\,kpc.

\subsection{\Cn\ \mgab}
\mgab\ was discovered by Robert H. McNaught (Coonabarabran, NSW, Australia) at magnitude 5.3 on 2020 July 15.590 UT (CBET 4811). It is also identified as Nova Reticuli 2020, and MGAB-V207, with Right Ascension of 03 58 29.55 and Declination $-$54 46 41.2 (J2000), its Galactic coordinates 265.397 $-$46.395. Following \citet{McLoughlin2021}, we take the date of discovery as 2020-07-08.1708, or JD 2459038.6708, henceforth \days{0}.

The lower panel of Figure \ref{fig:novae_light_curves} shows the AAVSO light curve of \mgab\ for the 125 days following eruption. \mgab\ exhibits a plateau-type light curve \citep{Strope2010} between \days{31} and \days{62}, which may relate to the plateau in the light curve of \cn\ V407 Cyg, interpreted by \citet{Hachisu2012} as a surviving accretion disc emerging out of the receding photosphere. The decline time of this \cn\ is $t_2 = 15$\,days during the initial decline (although this changes during the plateau event) \citep{ATEL14043}. A summary of the behaviour in \xrays\ and in gamma rays is given in \citet{McLoughlin2021}, and we refer the reader there for more detail. As discussed in that paper, the GAIA distance is $2.7\,^{+0.4}_{-0.3}\,\si{\kilo\parsec}$ \citep{Bailer-Jones2018}.  It was classified as a \HeN\ \cn\ \citep{ATEL13874}, while \citep{ATEL14048} suggest that this is an ONe white dwarf.

\subsection{A heterogenous sample}
The set of novae we examine includes one of each kind of the Tololo scheme for \cn\ spectra; one \feii, one \HeN, and one hybrid as detailed in Table \ref{tab:variety_of_novae}. The light curves (shown in Figure \ref{fig:novae_light_curves}) are thoroughly distinct, in their classifications, peculiar idiosyncrasies and timescales. Also, the underlying white dwarf compositions include both ONe and CO types. In Table \ref{tab:variety_of_novae}, we summarise the key details of three \cne.

\begin{table}
\caption{Properties of the classical novae. \newline 
\squarebrackets{a} \citet{Banerjee2015};
\squarebrackets{b} \citet{Pavana2020};
\squarebrackets{c} \citet{ATEL13874};
\squarebrackets{d} \citet{Aydi2020a};
\squarebrackets{e} \citet{McLoughlin2021};
\squarebrackets{f} \citet{ATEL14043};
\squarebrackets{g} \citet{Gehrz2018};
\squarebrackets{h} \citet{Sokolovsky2020};
\squarebrackets{i} \citet{ATEL14048};
\squarebrackets{j} \citet{Wee2020};
}
\label{tab:variety_of_novae}
\begin{tabular*}{\columnwidth}{lrrr}
\hline
                        &  V5668 Sgr                        & V906 Car                            & YZ Ret                         \\ \hline
Tololo class            & \feii\squarebrackets{a}           & Hybrid\squarebrackets{b}            & \HeN\squarebrackets{c}         \\
Light curve             & Dust-dip\squarebrackets{a}        & Jitter/Cusp\squarebrackets{d}       & Plateau\squarebrackets{e}      \\
Decline time $t_2$      &                                   & \days{44}\squarebrackets{d}         & \days{15}\squarebrackets{f}    \\
SSS on                  & \days{168}\squarebrackets{g}      &                                     &                                \\
SSS off                 & \days{240} to \days{340}\squarebrackets{g}&                             &                                \\
White dwarf type        & CO\squarebrackets{g}              &  CO\squarebrackets{h}               & ONe \squarebrackets{i}         \\
White dwarf mass        & $1.1\,M_{\odot}$\squarebrackets{g}& $0.71\,M_{\odot}$\squarebrackets{j} &                                \\
\hline
\end{tabular*}
\end{table}

\section{Observations} \label{sec:observations_novae}
The observational set-up and reduction pipeline for the data used in this paper are largely identical to the procedures described in our previous paper, \citet{McLoughlin2021}, so we here provide only a brief summary.

\subsection{Observatories}
The spectroscopic data in this paper is based on three major observational campaigns carried out by the \GJW\ telescopes, a system of 5 terrestrial observatories separated around the world by longitude to enable continuous monitoring of rapidly evolving astrophysical targets. The observatories house 0.5\,m telescopes, each with an \Aquila\ spectrograph which gave most of the observations and cover the wavelength range of approximately \angstrom{5800} to \angstrom{8400} with a resolution of $R\,{\sim}\,4000$. The majority of our spectra had exposure times of \exposure{100}, \exposure{300}, \exposure{1000}, \exposure{3000}. The shorter exposure times were employed to ensure that the \halpha\ line did not saturate to enable high-fidelity fitting of this complex, and longer exposures were used to give good signal-to-noise for the rest of the spectrum. The longer exposure time were necessitated as the \cne\ faded. By \days{1000}, all our \cne\ were typically observed with an exposure time of \exposure{3000}.

\subsection{Major campaigns}
The time-sampling we have on each of the three major campaigns discussed in this paper differ qualitatively, and as such the insights we gain from each differs accordingly.  We summarise each campaign in Table \ref{tab:observational_campaigns}.

\novasgr\ erupted 6 years ago, and we have data spanning most of the intervening time. This allows us to trace the different components over broad changes, and enables quantitative comparisons between early spectra and later analogues. We were particularly fortunate that this \cn\ erupted just after the system emerged from behind the sun, allowing for several months of consistent follow-up observations.

For \novacar, our campaign was focused on dense time-sampling for the first three weeks, with somewhat sparser observations beyond this. The strength of this campaign is that we were able to capture and precisely time (to within hundreds of seconds) the exact nature of certain early changes to the line profiles.

Our coverage of \mgab\ was a hybrid of these two extremes, with a virtually uninterrupted sequence of nightly spectra captured across the four months between \days{50} and \days{175}. This enabled us to `crack the code', and determine that a particular set of five \gaussian\ components (two pairs and an additional component) could satisfactorily explain the entire progression of its spectral changes over several months \citep{McLoughlin2021}. We used this model as a template to aid understanding of \novasgr\ and \novacar, and discuss the similarities and differences between these three examples of the varied field of objects that is \cne. 

\begin{table}
\caption{Data set summary for the three major observational campaigns}
\label{tab:observational_campaigns}
\begin{tabular*}{\columnwidth}{lrrr}
\hline
                        &  V5668 Sgr                             & V906 Car    & YZ Ret                              \\ \hline
First spectrum          & 2015-03-20                             & 2018-03-25  & 2020-07-15                          \\
Last spectrum           & 2020-10-01                             & 2020-12-28  & 2021-01-28                          \\
Discovery JD            & 2457097.134                            & 2458193.819 & 2459038.670                         \\
Aquila spectra          & 2792                                   & 3166        & 5226                                \\
LHires spectra          & 1                                      & 5           & 36                                  \\
Total exposures (h)     & 304                                    & 172         & 593                                 \\
\hline
\end{tabular*}
\end{table}

\subsection{Data reduction}
The data reduction pipeline used is called \codeendeavour, and it performs the necessary dark subtraction, flat-fielding and wavelength calibration to produce one-dimensional output spectra from the input two-dimensional CCD data. We then use a bespoke spectroscopic analysis tool called \poirot\ to perform the heliocentric correction and then carry out fitting, following the same protocols as those detailed in section 2.6 of \citet{McLoughlin2021}. The data are not photometrically calibrated, and so for visual consistency all the plots in this paper are normalised to unity. 

\section{Timing phase transitions for \novacar} \label{sec:novacar_timing_phases}
The existence of distinct phases has been known for some time, as mentioned in Section \ref{sec:introduction}. Here, with an intensive follow-up campaign for the first three weeks after \novacar\ erupted, we are able to quantify the duration of transitions between these phases, shedding light on the underlying physical mechanisms driving the observed changes in the optical spectroscopy.

\subsection{Timing the high-velocity absorption in \novacar} \label{sec:novacar_timing_shell}
\begin{figure}
	\includegraphics[width=\columnwidth]{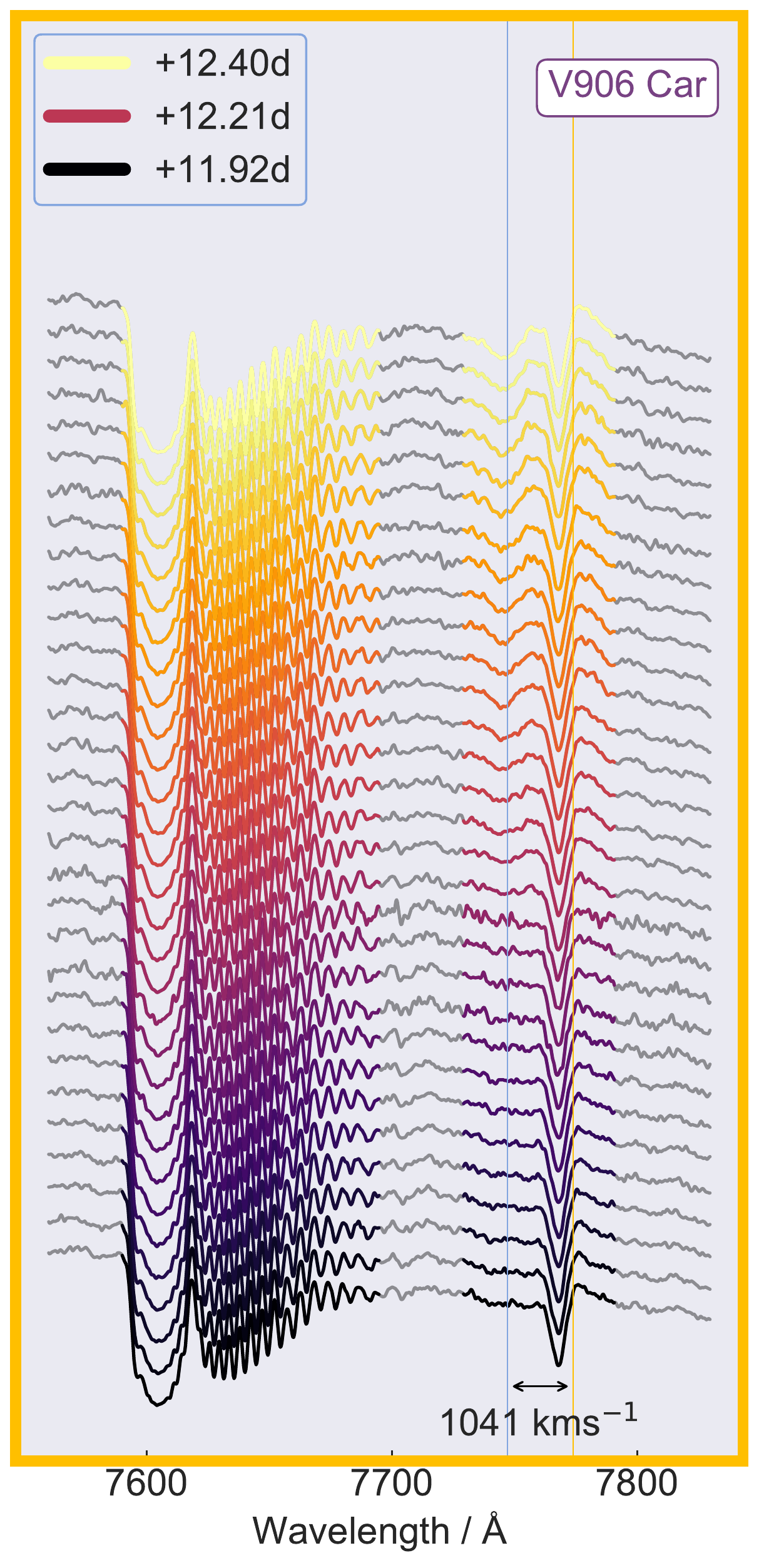}
    \caption{This series of spectra, with time increasing up the page, shows the development of a new high-velocity \tripleseven\ absorption line in \novacar, because of highly time-resolved sampling. The telluric-A band shown to the blue is included as an indicator of the robust wavelength calibration. The significant dates marked in the legend are the date of the last spectrum in the series, the date of the last spectrum with a non-detection of the high-velocity absorption component, and the date of the first spectrum in the series respectively. The thin vertical gold line corresponds to the rest wavelength for the \tripleseven\ line, while the pale blue line to the left indicates the wavelength of absorbing gas blue-shifted by \kms{1041}.}
    \label{fig:novacar_new_absorption}
\end{figure}

Figure \ref{fig:novacar_new_absorption} shows the early evolution of the [\ion{O}{i}] \angstrom{7774} complex alongside the telluric-A band (included to show stability of the wavelength calibration). Time runs from bottom to top, and the whole series lasts only half a day. In this time, the principal absorption component at \angstrom{7766} (\kms{300}) stays relatively steady, whilst a new and higher-velocity absorption complex appears at \angstrom{7743} (\kms{1200}), shown with a fiducial vertical line. The last non-detection of the secondary absorption is at \days{12.21}, and by \days{12.4} there is a clear detection. We are thus able to constrain the duration of this rapid phase transition to $< 5\,$hours; \citet{Aydi2020a} had noticed such a transition for \halpha\ and \hbeta\ to have occurred in their spectra taken a day apart.

\subsection{Timing the jet phase onset in \novacar} \label{sec:novacar_emission_onset}
While it has been well-understood for many years that \cne\ can quickly transition from single to multiple absorption, we believe these observations constitute the finest time-resolution yet for timing the transition.

\begin{figure*}
	\includegraphics[width=\textwidth]{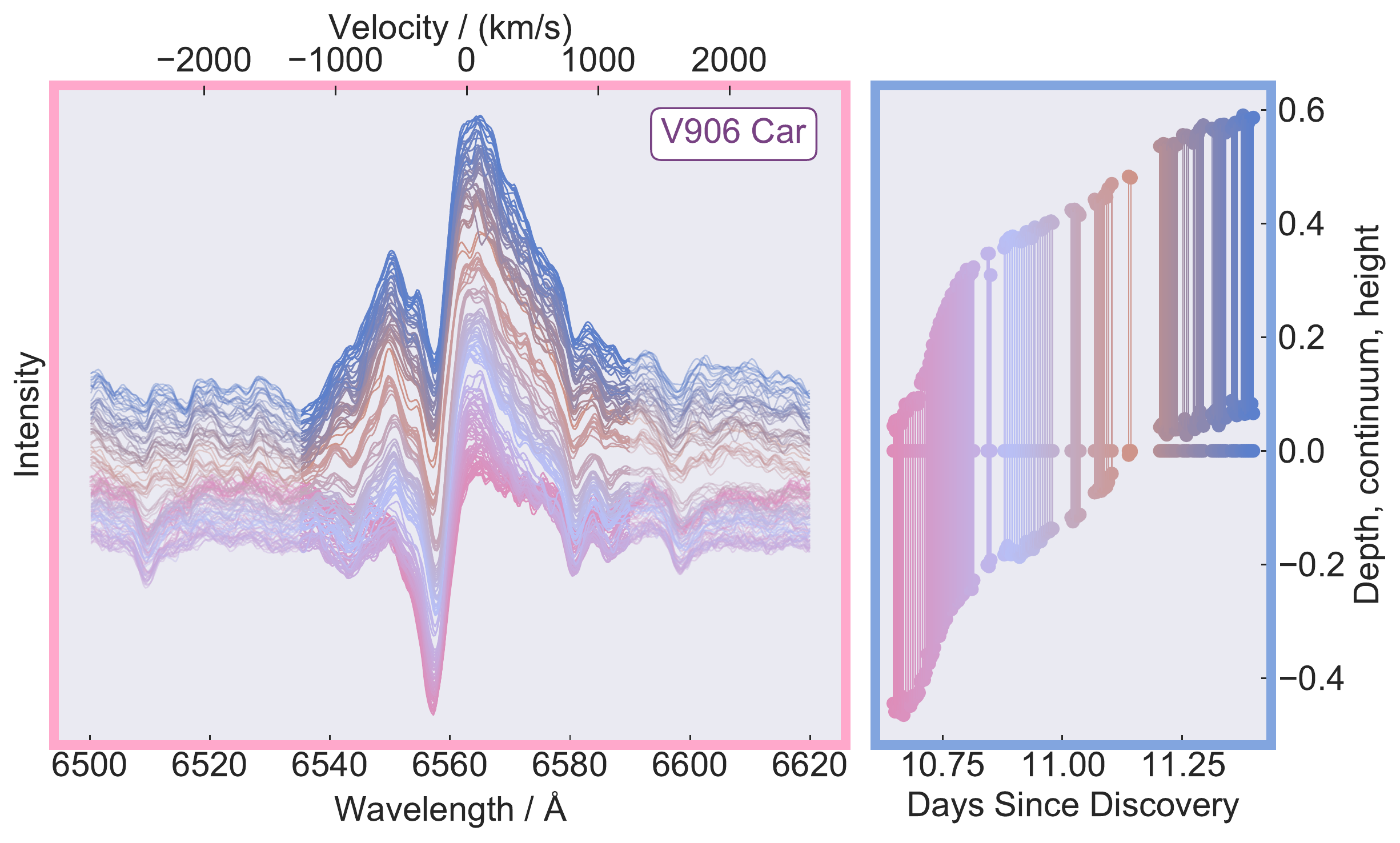}
    \caption{The left panel here shows a stacked series of the \halpha\ complex of \novacar, ranging from \days{10.6} to \days{11.3}, with the earlier spectra shown at the bottom in pink and the later observations at the top in blue. Each vertical line in the right panel represents one of these observations; the lower end of each vertical line represents the depth of the low-velocity absorption component, while the upper end is set to the height of the emission peak, with the continuum subtracted and accordingly set to zero. The latest observation has the absorption dot above the background level, because by this time the emission is so strong and broad that even the deep low-velocity absorption trough does not descend below the continuum.}
    \label{fig:novacar_eruption}
\end{figure*}

Figure \ref{fig:novacar_eruption} shows the sudden onset of significantly increased emission in \novacar. The left panel shows the spectra stacked with time increasing upwards, normalised from zero to one, while the right panel shows (with the corresponding colour for each spectrum) the depth of absorption and height of emission relative to that normalisation. Each vertical line in the right panel represents a single observation. The lower end of each vertical line represents the depth of the low-velocity absorption component, while the upper end is set to the height of the emission peak. This then shows that over the course of only half a day between \days{10.6} and \days{11.3}, there is a transition from predominantly absorption to predominantly emission. At the end of this sequence, the low-velocity absorption component is still present, but the minimum is now above the continuum level, due to the immense emission at that same wavelength.

The emission is hard to fit well while the \pcygni\ absorption is present. As explained in Section \ref{sec:jets_and_shell_ejecta}, \citet{McLoughlin2021} found that five emission \gaussian\ components accurately fitted the \halpha\ complex of \cn\ \mgab, and further showed that these five components split naturally into a fast-moving pair, a slow-moving pair, and a single additional central emission line. The pairs were then identified as representing jets and an accretion disc respectively, while the additional emission, which was less dynamically significant, and presumed to be emanating from the slow-moving spherical ejecta. With the diminishing of the absorption systems in \novacar\ by \days{30}, we obtain an uncontaminated view of the emission profile, revealing an emission complex that is well fit by this jets and accretion disc model. We therefore infer that, since the qualitative shape of the emission does not significantly change between \days{10} and \days{30}, the fresh emission shown in Figure \ref{fig:novacar_eruption} is most likely due to the jets and accretion disc just as in \mgab\ in \citet{McLoughlin2021}.

There are two plausible explanations for this sudden onset of \halpha\ emission in the context of jets in \novacar. One is that it takes a certain amount of time for the recently ejected material in the immediate vicinity of the eruption to be processed down the accretion disc and then launched as jets. In this scenario, one would expect to see the accretion disc emerging prior to the jets being formed; in the right-hand panel of Figure \ref{fig:novacar_eruption}, we note a distinct change in the gradient of growth of emission occurring at \days{10.8} and speculate that this may mark the onset of jets.

The second possible conjecture is that the disc, fuelled by the recently ejected material, immediately forms jets which, initially buried in the heavily absorbing shell-like ejecta, emerge at this epoch and become spectrally significant.

\begin{figure*}
	\includegraphics[width=\textwidth]{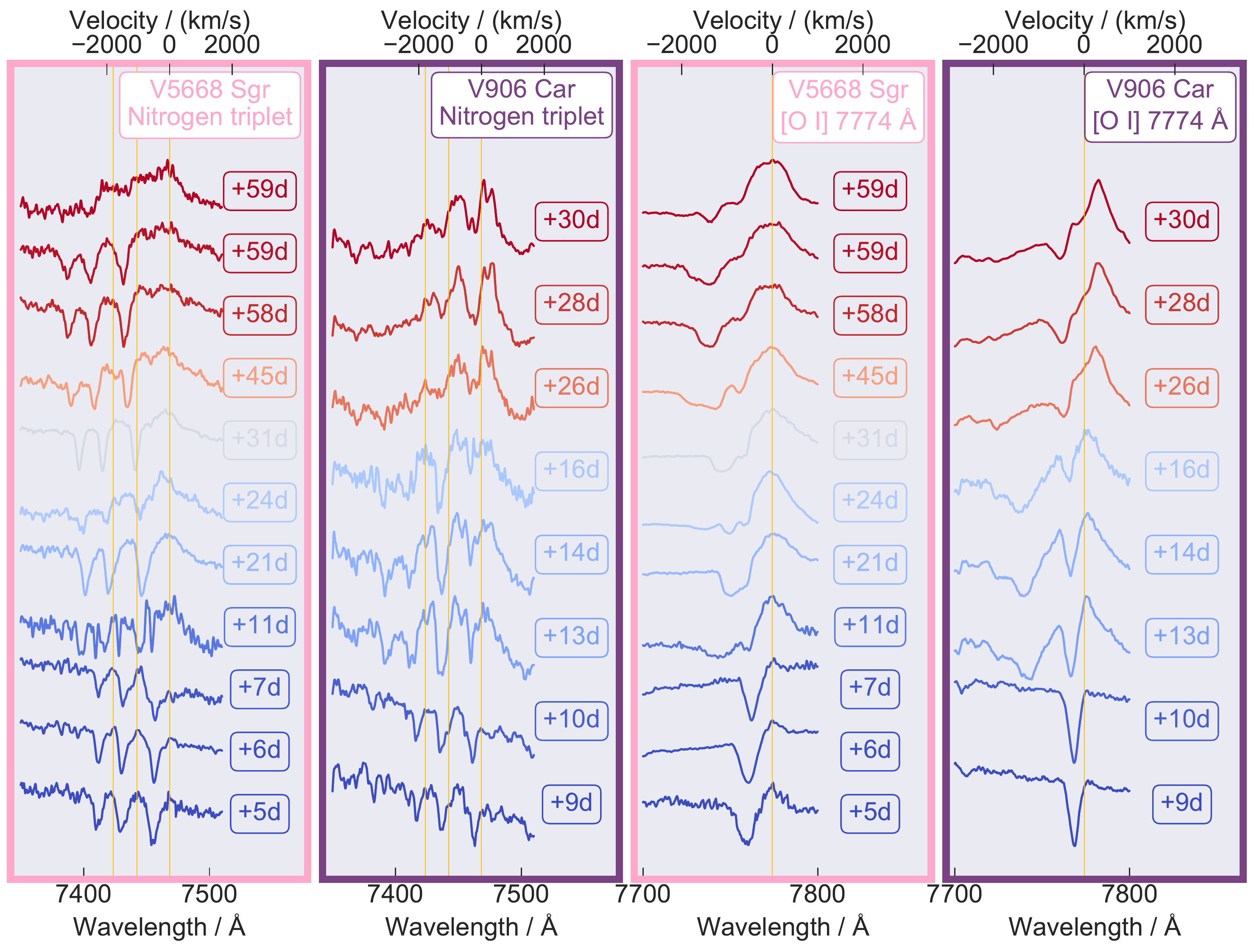}
    \caption{This figure compares the profiles of the nitrogen triplet (\angstrom{7423.641}, \angstrom{7442.298}, and \angstrom{7468.312}) and \tripleseven\ between the two \cne, \novasgr, and \novacar. The left two panels show the nitrogen complex in the two novae, while the right two panels show the oxygen progression. For \novasgr, we plot two separate spectra taken hours apart at \days{59}, which capture the disappearance of the nitrogen triplet - an event with no analogue in oxygen. The emission complex shown at the lower end of the wavelength range of \novacar\ is not part of the oxygen complex, but a circumbinary disc signature, which was the subject of a previous paper \citep{McLoughlin2020}.}
    \label{fig:novae_nitrogen_progression}
\end{figure*}

\subsection{Evolution of nitrogen and oxygen in \novacar} \label{sec:novacar_nitrogen_triplet}
The second panel in Figure \ref{fig:novae_nitrogen_progression} displays the evolution of the nitrogen triplet (\angstrom{7423.641}, \angstrom{7442.298}, and \angstrom{7468.312}) as it moves from absorption through to emission. From \days{13} and beyond, we note hints of splitting in the peaks of some of these lines. Although this probes the limit of sensitivity given the signal-to-noise in this faint complex, splittings of \about{\kms{200}} are suggested. These spectra also serve as a reference for the discussion of sudden changes in \novasgr\ in Section \ref{sec:novasgr_nitrogen_triplet}. 

In the right panel of Figure \ref{fig:novae_nitrogen_progression}, we show the progression of the \tripleseven\ complex of \novacar. Perhaps more clearly than in the equivalent nitrogen panel, the oxygen low-velocity absorption line moves to increasingly blue-shifted wavelengths over time. We then see the commencement of emission between \days{10} and \days{13}, which mimics the behaviour shown for the \halpha\ complex in Figure \ref{fig:novacar_eruption}.

\section{Timing phase transitions for \novasgr} \label{sec:novasgr_timing_phases}
Figure \ref{fig:novasgr_life_story} shows the progression of the \halpha\ complex of \novasgr\ through the several phases, discussed in this section. The three distinct phases that we focus on each approximately span a year in this \cn; in the following sections, we consider each in turn. We then comment on the rapidly evolving nitrogen triplet, and contrast this with that observed in \novacar\ in Section \ref{sec:novacar_nitrogen_triplet}.

\subsection{\novasgr: Initial jets and accretion disc}
\begin{figure*}
	\includegraphics[width=\textwidth]{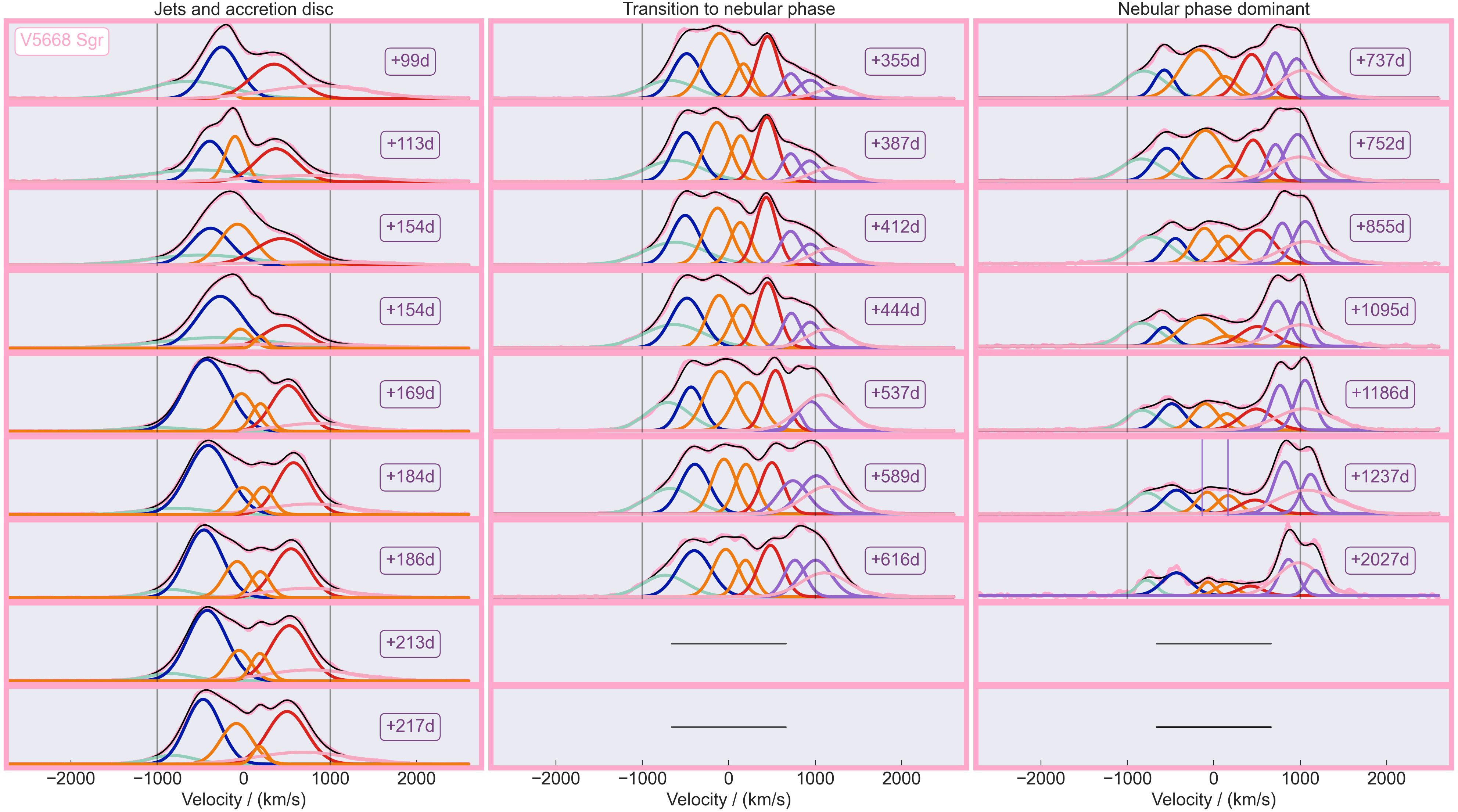}
    \caption{This figure shows an extensive sequence of \gaussian\ fits to the H-alpha complex of \novasgr, from \days{99} to \days{2027}, by which time the spectrum is dominated by nebular nitrogen emission. The first column shows roughly the first year, when a simple jets and accretion disc model suffices to explain the profiles - the blue-shifted and red-shifted jet components are shown in turquoise and pink respectively, while the double-peaked signature of the accretion disc is represented by red and blue \gaussians. Following \citet{McLoughlin2021}, there is an additional orange component most probably associated with an ejected shell. We show two separate models for \days{154}, since from this epoch it becomes necessary to include a second additional component to explain the data. The second column shows the \cn\  during its transition to the nebular phase. Here, the same jets and accretion disc model remains valid, but requires two additional components (shown in purple) which are consistent with being due to \nitrogensixfiveeightfour. The third column shows a continuation of this pattern during the subsequent years, with \nitrogensixfiveeightfour\ becoming the dominant emitter in this region. For the panel at \days{1237}, we add two purple vertical lines to the plot, which represent the radial velocities of the two nitrogen \gaussian\ components shown in purple, showing that these nebular lines are consistent with being from a dynamically similar parcel of gas as the shell (orange components) of H-alpha.}
    \label{fig:novasgr_life_story}
\end{figure*}

Here we apply the same model developed in \citet{McLoughlin2021}, and discussed in Section \ref{sec:jets_and_shell_ejecta}, consisting of a fast-moving pair of emission lines, a slow-moving pair, and additional central \gaussians, to the \halpha\ complex of \novasgr.

The fits shown for the first spectra in the first column of Figure \ref{fig:novasgr_life_story} show a remarkably similar set of \gaussian\ components to those observed in \mgab\ in \citet{McLoughlin2021}. In all cases, we removed spectra that were in any way compromised, for example by the onset of clouds or by saturation in the \halpha\ complex. The separation of the fast-moving pair of \gaussians\ is too fast to be plausible as material orbiting the white dwarf, yet far enough away to be cool enough to radiate Balmer \halpha\ emission (i.e.\ given the continued nuclear burning at the surface and temperatures in excess of $\num{e8}$\,K). We therefore infer that these spectra of \novasgr\ also represent anti-parallel jets (shown in turquoise and pink), and an accretion disc (blue and red), with an additional component (orange) perhaps formed in a shell-like outflow. Initially, this additional component does not account for a significant proportion of the flux relative to the far stronger jet and disc signatures. However, by \days{113}, the disc has begun to weaken in emission, and the jets have declined slightly too. 

By \days{154}, we begin to require two additional \gaussian\ components (other than the jets and disc) to fit the profiles well. Figure \ref{fig:novasgr_life_story} shows \days{154} twice: once with and once without an extra component to show its inclusion is necessary to obtain a close fit. The close similarity between these two additional components of \halpha\ and the nebular nitrogen lines which appear later (see Section \ref{section:novasgr_nebular_nitrogen_phase}) suggests that these are formed in the ejected shell, which is supported by \nitrogensixfiveeightfour\ rest-frame filtered imaging observations reported by \citet{Slavin1995} for several old \cn\ remnants.

\subsection{\novasgr: Transition to nebular phase} \label{section:novasgr_nebular_nitrogen_phase}
The middle column of Figure \ref{fig:novasgr_life_story} approximately spans the second year after eruption. During this time, additional lines appear to the red of the rest wavelength. We use the latest time spectra to get an accurate initial guess for the central wavelength of the extra components, and find that they are consistent with being a double-peaked [\ion{N}{ii}] \angstrom{6584} line. These spectra are well fit by the same set of six \gaussian\ components as those fitting the spectrum at \days{154}, but require the addition of two nitrogen lines (\nitrogensixfiveeightfour). With the goal of using the \halpha\ complex alone to determine the phase of a \cn, this change would provide a straightforward marker for the onset of the nebular phase. At this point, there is still comparable flux coming from \halpha, and so there is not a distinct moment at which the nebular lines begin to dominate. 

\subsection{\novasgr: Nebular phase dominant}
\begin{figure}
	\includegraphics[width=\columnwidth]{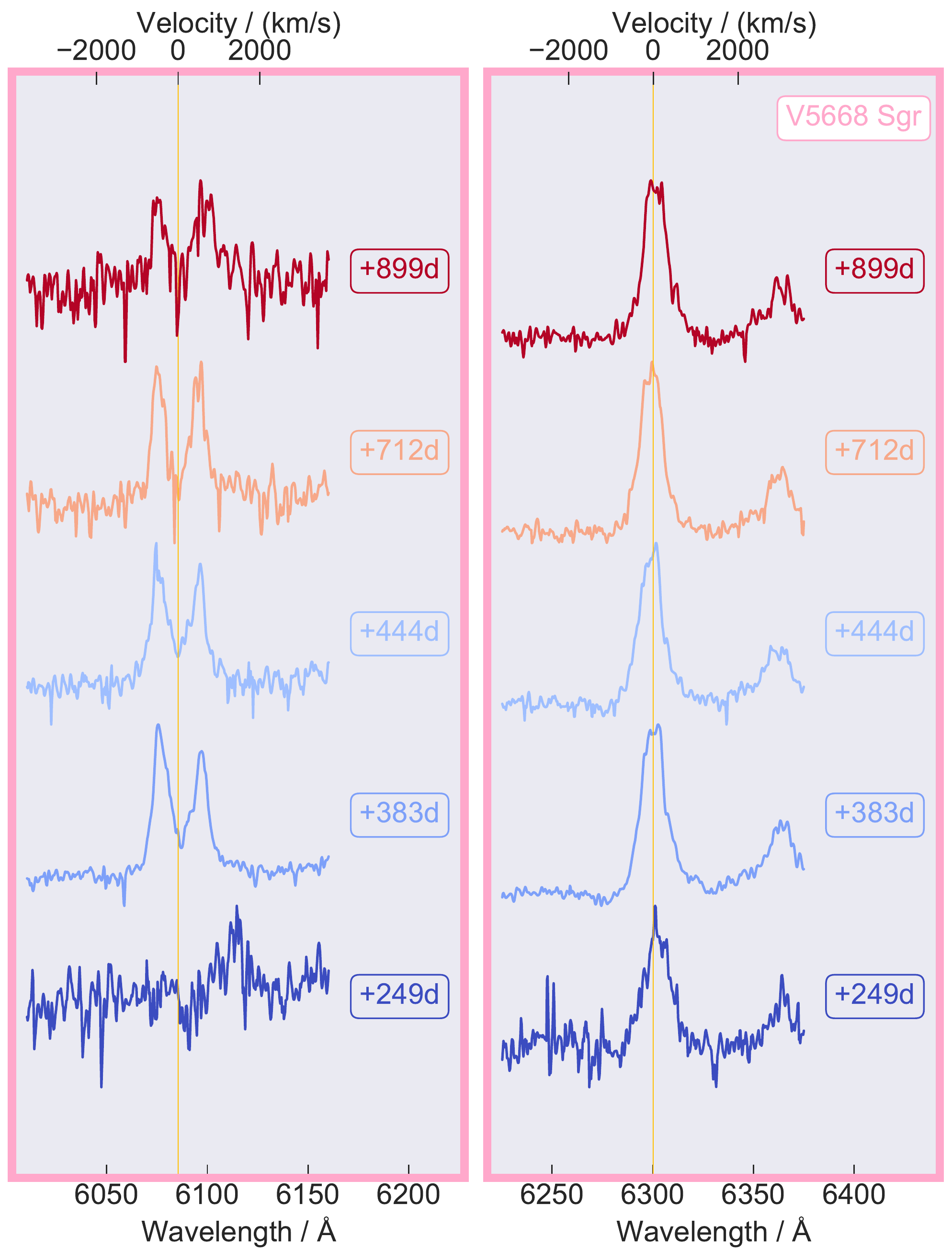}
    \caption{The left panel shows a double-peaked profile of the high-excitation coronal line of \fesixtyeightyfive\ in \novasgr, from when it first appears in our spectra around one year after eruption, to the last detection at \days{899}. The right plot shows the \oxygensixtythreehundred\ line, which is present throughout our spectroscopic monitoring campaign of \novasgr. }
    \label{fig:novasgr_nebular_fe_iron}
\end{figure}

In the third column of Figure \ref{fig:novasgr_life_story}, the nebular phase begins to dominate. By \days{750} the [\ion{N}{ii}] \angstrom{6584} profile gives the strongest emission in this wavelength range, and by \days{1100}, nitrogen is totally dominant. \citet{Gehrz2018} identify the presence of the high-excitation forbidden coronal line of \fesixtyeightyfive. Figure \ref{fig:novasgr_nebular_fe_iron} demonstrates the arrival and departure times of the \fesixtyeightyfive\ line in our data; the \oxygensixtythreehundred\ is present throughout the existence and demise of the \fesixtyeightyfive\ line. The \fesixtyeightyfive\ emission is contemporaneous with the nitrogen lines shown in Figure \ref{fig:novasgr_life_story} (in purple), which begin to emit strongly after approximately one year since detonation. Unlike these nitrogen lines, which establish themselves as the strongest lines in our optical spectra after \days{800}, the \fesixtyeightyfive\ line dwindles in comparison to \halpha\ at late times. 

The [\ion{N}{ii}] \angstrom{6584} lines have the same radial velocity as the two additional components of \halpha, shown in Figure \ref{fig:novasgr_life_story} by two purple fiducial vertical markers. The nitrogen only becomes important compared to \halpha\ at late times once the density has decreased sufficiently, since it is a forbidden transition; by this time, the strength of the \halpha\ equivalent pair of lines has diminished.

\subsection{\novasgr: evolution of nitrogen and oxygen} \label{sec:novasgr_nitrogen_triplet}
The left panel of Figure \ref{fig:novae_nitrogen_progression} shows the progression of the nitrogen triplet (\angstrom{7423.641}, \angstrom{7442.298}, and \angstrom{7468.312}) in \novasgr. What is fascinating about this is the suddenness of the disappearance of absorption lines between two spectra both taken on \days{59}. Using our time-resolved spectroscopy, we can put an upper bound on the duration of this phase transition of $t <= 17\,$h. The absorption was not present in any of the subsequent spectra. 

There is a trend in the radial velocity of nitrogen absorption lines, which moves to higher speeds as time progresses between \days{7} (\kms{-480}) to \days{59} (\kms{-1440}). In the very early spectra taken before \days{7} though, it decreases in radial speed, which we discuss in more detail below. Our spectrum taken at \days{11} shows at least two distinct features, a narrow low-velocity component at \kms{540} and a broader high-velocity complex between \kms{802} and \kms{1000}. It is therefore likely the dominant line in our later spectra is related to the high-velocity component, rather than a drastically shifted low-velocity component.

The third panel of Figure \ref{fig:novae_nitrogen_progression} follows the \tripleseven\ line through a series of changes. It starts off at \days{5} and \days{6} with a single low-velocity \pcygni\ absorption feature with a speed of \kms{580} along the line-of-sight. This then slows, reaching a minimum speed of \kms{420} on \days{7}, before progressively increasing in blue-shift up to a maximum of \kms{730} on \days{45}. This pattern of slowing before continuing to advance to the blue, which is also present in our data of this nitrogen complex, has previously been recorded in \novacar\ by \citet{Aydi2020}, and is only apparent when spectral data are captured close to or before optical maximum. \novasgr\ has many large-scale jitters in the early light curve (Figure \ref{fig:novae_light_curves}), and this is probably why we are able to see this early `retrograde' pattern. 

From \days{11}, the low-velocity component is joined by a high-velocity system, with a wide range of absorption velocities. This high-velocity absorption system of \novasgr\ was the subject of a thorough study with high-resolution data \citep{Tajitsu2016a}, which studied a snapshot of the Balmer lines of hydrogen and other metal lines, and measured the precise wavelengths of various narrow components within the complex. 

\section{Overview: A dynamical model for nova phases} \label{sec:novae_overview}
\begin{figure}
	\includegraphics[width=\columnwidth]{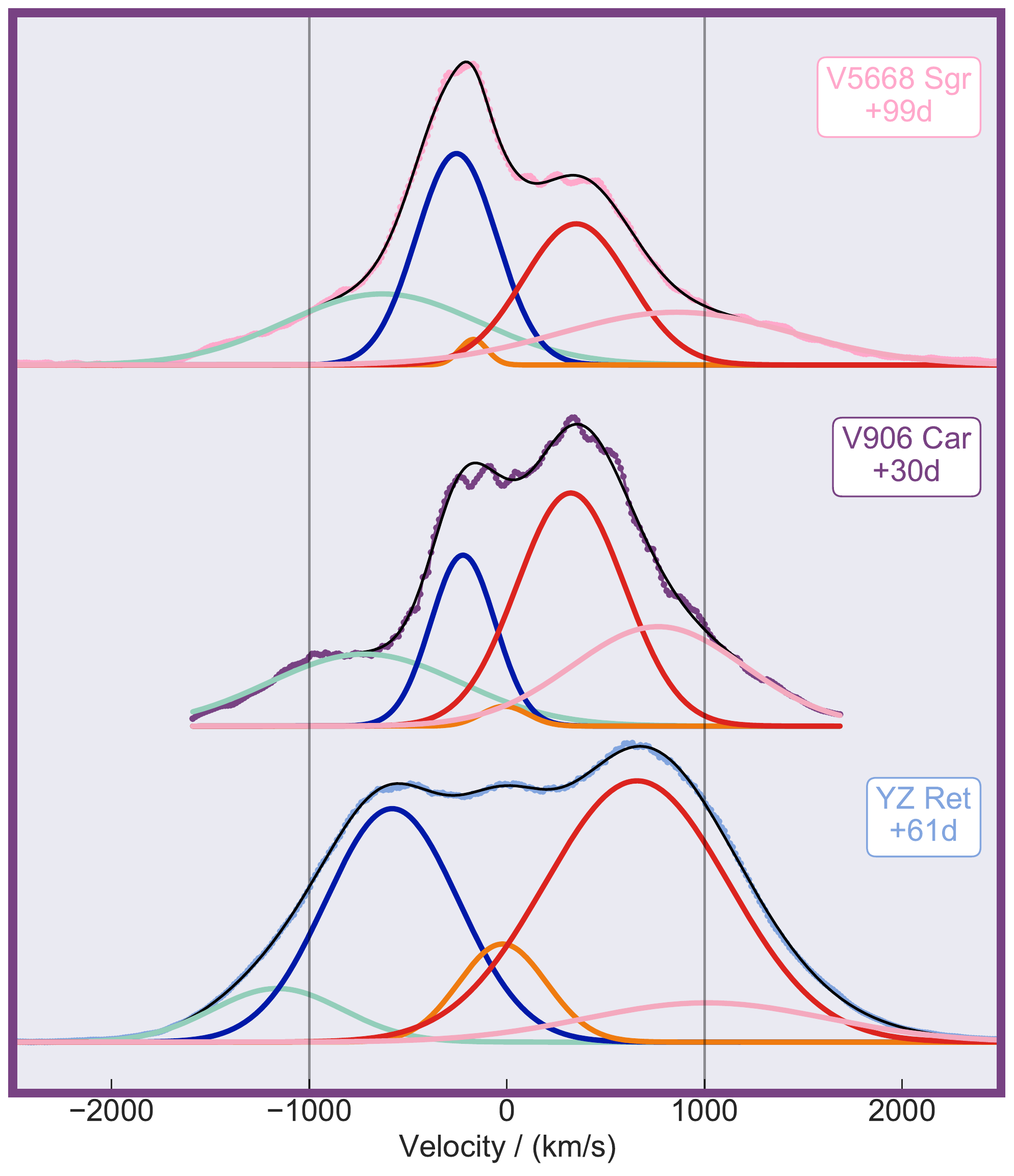}

    \caption{Here we show our five-component \gaussian\ model fits to the H-alpha complex of three separate and quite different \cne. The lower spectrum is of \mgab\ at \days{61}, the middle is of \novacar\ at \days{30}, and the top is of \novasgr\ at \days{99}. These \cne\ had very different light curves (shown in Figure \ref{fig:novae_light_curves}), yet all are well fit by two pairs of lines and an additional component. The colour scheme and interpretation of lines follows Figure \ref{fig:novasgr_life_story}. There are two vertical fiducial markers, at $\pm$\kms{1000}.}
    \label{fig:three_novae}
\end{figure}

Figure \ref{fig:three_novae} shows the striking similarity between three seemingly very different eruptions. For each of these systems, we have coloured the putative jet components in turquoise and pink, with the accretion disc in blue and red. The jet speeds differ substantially, although we cannot here separate the effects of inclination angle from the actual rest frame jet-launch speed.

\subsection{Onset of jets and accretion disc in \novacar} \label{sec:novacar_jets_and_disc}
\begin{figure*}
	\includegraphics[width=\textwidth]{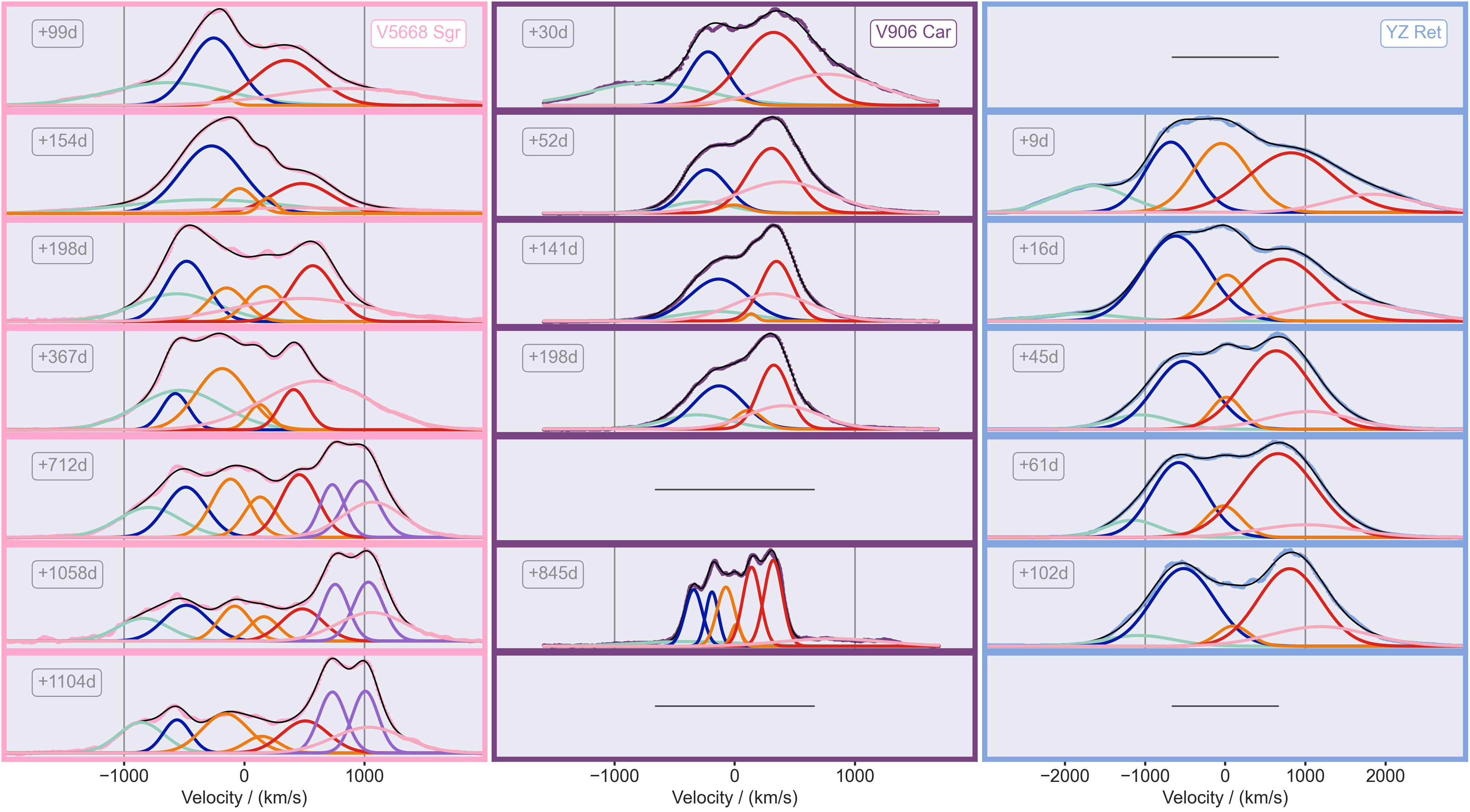}
    \caption{Here we demonstrate the applicability of the same jets and accretion disc model used to successfully describe the H-alpha evolution of \cn\ \mgab\ \citep{McLoughlin2021} to two additional examples --- \novasgr\ in the left column, and \novacar\ in the right column. \novasgr\ shows two additional \gaussian\ components (purple) which are due to \nitrogensixfiveeightfour. Since the campaign for \novacar\ was primarily focused on the first 3 weeks after eruption during the multiple absorption component phase, our data set for this target has a substantial break before the final spectrum. Despite not being able to fully track the components through a time series, the \days{845} spectrum of \novacar\ shows hints of splitting occurring in the jet lines, coloured turquoise and pink.}
    \label{fig:jet_gallery} 
\end{figure*}

Our observational campaign for \novacar\ focused keenly on the first three weeks, which made it possible to capture several key early transitions in the spectroscopic development of this \cn. However, as shown in panels two and four of Figure \ref{fig:novae_nitrogen_progression}, the emerging emission spectrum is interspersed with several highly variable absorption systems during the first month following eruption. In Figure \ref{fig:jet_gallery} therefore, we begin fitting the emission complex for \novacar\ (middle column) from \days{30}, by which time the complex is well-fit by the standard five emission \gaussians\ model we previously identified as jets and accretion disc for \mgab\ and \pnvj\ in \citet{McLoughlin2021}. This fit remains accurate at least until \days{198}. Following Section \ref{sec:novacar_emission_onset} where we discuss the onset of substantial emission in \novacar, it is reasonable to assume that the emission complex that emerges suddenly around \days{10} would also, in the absence of complex absorption systems, be well-fit by the same model as the one which remains in place from \days{30} to \days{200}. We therefore feel confident placing \novacar\ in the collection of \cne\ which display jets and an accretion disc in their \halpha\ profiles, and believe that this is the emission responsible for the initial rise in \halpha\ flux from \days{10.6} to \days{11.3}.

\subsection{An updated prescription for \cn\ optical evolution}
There is a small but growing sample of \cne\ which show jets and an accretion disc in their \halpha\ spectra, which are the two examples we have presented in this paper (\novacar, \novasgr), together with the two in \citet{McLoughlin2021} (\mgab\ and \pnvj, in their figures 3 \& 8 respectively). We therefore suggest an update to the traditional physical model underpinned by optical observations. At first, there is a \pcygni\ spectrum, with a relatively low velocity absorption component on the order of hundreds of \kms{} (the LVC), most likely associated with a slowly-ejected shell. This is promptly followed by the arrival of secondary absorption, with far higher radial velocities on the order of thousands of \kms{} (the HVC), which often consists of multiple separate components. Very shortly after this, emission features originating from material at the base of jets and from the accretion disc emerge, taking only a few hours to push the minimum of the LVC absorption trough above the continuum level. The jets and the accretion disc appear to be precessing in \mgab, and similarly systematic follow-up of other \cne\ may well yield similar results. At late times, nebular lines of forbidden transitions become strong, a process through which [\ion{N}{ii}] \angstrom{6584} took approximately a year to compete with \halpha\ in \novasgr, and approximately another year to become the dominant source of emission in our optical spectra. 

\subsection{Jets, not clumps, responsible for H-alpha spectra}
What is most remarkable about the fits shown in Figure \ref{fig:jet_gallery} is how well this model accounts for variability in the line profiles. It is not warranted to claim that mere ``clumpiness'' can account for all this complexity. While we do of course agree that the ejecta does form clumps of some description, as suggested by imaging of \cn\ remnants (such as the \alma\ imaging of \novasgr\ by \citet{Diaz2018}, though detailed conclusions are hindered by a lack of short UV-spacings), we believe that our data set shows that it is not necessary to invoke clumps in order to explain the complex profiles. Instead, a dynamical model consisting of jets, an accretion disc and a simple shell outflow is sufficient to explain the spectra from the early fading of absorption systems until the late nebular phase. Of course, the [\ion{N}{ii}] lines need to be considered also, but they appear at late times, following the same dynamical prescription as the \halpha\ shell, and as such are easily understood.

\subsection{Implications for interstellar enrichment}
Another interesting consequence of the developments to the jet model for nova outflows is that it informs our understanding of the interstellar enrichment capabilities of \cne. Considering the ejection to take place from near the surface of the white dwarf, where the escape velocity is approximately given by $V^{2}_{\text{esc}} \sim G M_{\odot} / R_{\text{WD}}$ which yields a value of around \kms{4500}, in excess of the shell ejection speeds observed in the majority of systems which thus implies that these do not contribute significantly to the enrichment of the interstellar medium (ISM). In the jet paradigm however, material is not launched from the white dwarf surface itself but is instead launched from the accretion disc. Jet models based on young stellar objects show the jets to be launched from the co-rotation radius, where the Keplerian accretion disc angular speed matches that of the white dwarf spin, and these indicate that the jet speed is proportional to the Keplerian speed with a small, order-unity constant \citep{Tzeferacos2013,Shu1994}. For a \cn\ with a canonical mass of \solarmasses{1} and a spin period of \exposure{60} following the \xray\ evidence hinted at by \citet{Osborne2015}, this gives $v = (GM\omega)^{1/3} \approx $ \kms{2400} as the Keplerian velocity of the disc at the co-rotation radius, and thus an approximate estimate for the jet speed in the frame of the eruption (here $G$ is the gravitational constant, $M$ is the enclosed mass and $\omega$ is the angular frequency). This speed is comparable with the jet launch speeds observed in \mgab\ and \pnvj, of \kms{2000} and \kms{4500} respectively, given the uncertainty introduced by poorly constrained inclination angles.

The escape velocity from the co-rotation radius ($R_{\text{co}}$) of a canonical \cn\ is given by $V^{2}_{\text{esc}} \sim G M_{\odot} / R_{\text{co}} \approx$ \kms{3600}, which is greatly reduced compared to that at the surface of the accretor and, crucially, is less than the likely jet launch speed for the fastest jets we observe (especially since observed radial speeds are lower than the nova-frame speeds by expected projection effects). The consequence of this is that \cne\ with fast jets may therefore play a role in enrichment of the ISM. In the case of \novasgr, the jets appear to last in excess of five years; in this time, the initial ejecta will have reached a distance of at least 1000\,AU. The stellar mass density close to the Galactic centre is approximately \solarmasses{\num{e8}}\,pc$^{-3}$, which, assuming typical stellar masses of \solarmasses{10}, suggests an average distance between stars of $955\,$AU. Therefore it seems likely that the jet ejecta of \cne\ can indeed penetrate the surrounding ISM, increasing its enthalpy and temperature. With \about{100} \cne\ per year in the Milky Way, and a typical mass ejection of around \solarmasses{\num{e-4}} per nova \citep{Yaron2005}, the cumulative enrichment effect of \cne\ on the Galactic ISM may be on the order of \solarmasses{\num{e7}} every billion years.

\subsection{Prediction for spatially resolved spectroscopy}
The sprouting of the lines of \nitrogensixfiveeightfour\ (shown in purple in Figure \ref{fig:novasgr_life_story}) around at the anniversary of the explosion, and their resemblance to the central lines of \halpha, is a reminder that detailed spectrally-resolved imaging of \cne\ a few years after eruption would likely reveal structures that could confirm or refute the jets model. In the case of the B[e] star studied in \citet{Mehner2016}, an 8\,m class telescope (VLT) together with a modern IFU (MUSE) revealed a slender jet in the \nitrogensixfiveeightfour\ line (their figure 8). With sensitive, high-angular resolution observations, it would be very instructive to see nitrogen-free \halpha\ imaging, and to see which shells and which elongated outflows appear in emission lines besides hydrogen.

\section{Conclusions} \label{sec:conclusions}
The designation of phases for the optical spectroscopic progression of \cne\ has typically focused on relatively early-time absorption lines to elucidate what is happening specifically along the line-of-sight to the eruption \citep{Mclaughlin1956}. However, a full understanding of the nature and evolution of the \cn\ phenomenon necessitates sustained time-resolved studies beyond this regime. Traditionally, the \halpha\ emission complex has been deemed too contaminated to be straightforwardly fit and understood. In this paper, we extend the sample of two \cne\ in \citet{McLoughlin2021} to include two further \cne, and show that a jets and accretion disc model for understanding the \halpha\ emission is perhaps more widely applicable to \cne\ than was previously imagined. With the sample size growing, we propose a simple addition to the standard sequence of events and phases in optical spectra and the way of understanding them. The variety of the \cne\ now in the sample is encouraging, since it suggests that the jets and accretion disc model may be a feature of \cne\ in general, rather than just restricted to one type.

We provide tight constraints on the timing of transitions between these classic phases, which lends support to the conjecture that jets and an accretion disc are key to understanding the optical emission of \cne\ in the first thousand days. 

We further speculate that thinking of the emission in terms of jets launched by an accretion disc rather than from the white dwarf surface could alter the relevant escape velocity, and may therefore allow for enhanced enrichment of the interstellar medium compared to what was previously thought possible from \cne. 

Beyond the spectroscopic evidence we have presented for these novae, independent support for our conjecture would come from a spatially-resolved spectroscopic image of these targets in a few years time, when the angular sizes would have grown sufficiently large to be well-matched to modern IFUs.

Such detailed evolving spectral behaviour now warrants development of a detailed underlying physical model. This would need to explain how the accretion process launches jets in the immediate aftermath of classical nova outbursts as well as the detailed observational features of the spectroscopic signals we have presented here and in \citet{McLoughlin2021}. These include (a) the suddenness of the onset of emission (b) the coupled precession of the jet and accretion disc radial velocities (c) the persistence of jet emission for at least several years following the eruption.

It is possible that the pursuit of a physical model elucidating the launching of jets after the nova eruption in a system that, prior to the explosion did not show jets, may yield important insights on the astrophysical jet phenomenon as a whole. This may inform studies of systems which show similar behaviour, albeit on a variety of scales with a dynamical range of many orders of magnitude (including young stellar objects, microquasars and supermassive black holes).

\section*{Acknowledgements}
A great many organisations and individuals have contributed to the success of the Global Jet Watch observatories and these are listed on www.GlobalJetWatch.net but we particularly thank the University of Oxford and the Australian Astronomical Observatory. DM thanks the STFC for a doctoral studentship, and Oriel College, Oxford, for a graduate scholarship.  We acknowledge with thanks the variable star observations from the AAVSO International Database contributed by observers worldwide and used in this research.

\section*{Data availability}
Light curve data for all our targets is made available by the AAVSO, at \href{aavso.org}{aavso.org}. The data underlying this article were provided by the \GJW\footnote{\href{www.GlobalJetWatch.net}{www.GlobalJetWatch.net}} by permission. Data will be shared on request to the corresponding author with permission of the \GJW.




\bibliographystyle{mnras}
\bibliography{references} 







\bsp	
\label{lastpage}
\end{document}